# Clustering based Privacy Preserving of Big Data using Fuzzification and Anonymization Operation

Saira Khan[1], Khalid Iqbal[2], Safi Faizullah[3], Muhammad Fahad[4], Jawad Ali[5], Waqas Ahmed[6]

Department Computer Science, COMSATS University Islamabad, Attock Campus, Pakistan[1,2,4]
Faculty of Computer and Information System, Islamic University of Madinah, Saudi Arabia[3]
Malaysian Institute of information Technology (MIIT), University Kuala Lumpur, Kuala Lumpur, Malaysia[5]
UniKL Business School (UBIS), University Kuala Lumpur, Kuala Lumpur, Malaysia[6]

*Abstract*—Big Data is used by data miner for analysis purpose which may contain sensitive information. During the procedures it raises certain privacy challenges for researchers. The existing privacy preserving methods use different algorithms that results into limitation of data reconstruction while securing the sensitive data. This paper presents a clustering based privacy preservation probabilistic model of big data to secure sensitive information..model to attain minimum perturbation and maximum privacy. In our model, sensitive information is secured after identifying the sensitive data from data clusters to modify or generalize it.The resulting dataset is analysed to calculate the accuracy level of our model in terms of hidden data, lossed data as result of reconstruction. Extensive experiements are carried out in order to demonstrate the results of our proposed model. Clustering based Privacy preservation of individual data in big data with minimum perturbation and successful reconstruction highlights the significance of our model in addition to the use of standard performance evaluation measures.

## I. INTRODUCTION

The big data refer to the massive amount of structured and unstructured data with an increase of 2.5 ExaBytes per day. This rapid increase in data volume is due to web-services [1], mobile data [2, 3], health care data [4], GPS signals, Youtube, digital cameras and file hosting websites, and social media like facebook and twitter [5]. Generally, Big data can be classified into Volume, Velocity and Variety [5, 7]. First, Volume is the amount of data larger than tera bytes and peta bytes. Second, Velocity represents the speed of in and out, share and seized of data. Lastly, Variety means the explosion of new data types from social sites, mobile computing and machine devices. Big Data is used to bring improvements in businesses and society known as Big Data Analytics [6]. However, the combination of personal and external data increases the vulnerability of sensitive attributes. Prevention of sensitive information from disclosure is known as Big data privacy. The causes of privacy violation are the massive volume of data, management, storage, manipulation and data analytics [8]. Big data privacy can either be preserved by design or by rules and regulation as shown in Fig. 1.

Sanitization is common technique to remove the sensitive attribute in order to secure the quasi and sensitive information while preserving privacy of big data. The removal of sensitive and quasi attributes may lead to the incorrect mining results in different fields i.e., Healthcare and Bank Datasets [8]. The sensitive attributes in the datasets can either be numeric or categorical such as patient name, zip code, DOB, gender or sex, salary, balance, disease, race and Phone Numbers. These attributes threaten the individual privacy. To secure the individual privacy of structured data, anonymization algorithm is presented by Mohammad et al. [9] that probabilistically generalizes the data by adding noise to guarantee differential privacy. The advantage of generalization is to build a decision tree effectively. However, generalization is ineffective in re-construction of attribute. Sanitization lowers the re-construction ability of sensitive data over large scale [8]. A framework is proposed to conceptualize and measure service quality for internet of things. This study establishes the IoT-SERVQUAL model with four dimensions (i.e., Privacy, Functionality, Efficiency and Tangibility) of multiple service quality models and security issues[29].

On the other hand, Abitha et. Al [11] used min-max normalization, fuzzy logic, rail fence and map range to preserve the privacy of individual by masking the sensitive attributes. The use of fuzzy logic reduced the number of iterations and processing time that enhanced the clustering efficiency. However, S-shape membership function presented better results than the traditional data masking algorithms. In addition, Min-max normalization, rail-fence method and map range can only use for specific ranges in a given data, categorical data and numeric data respectively.

In this paper, a probabilistic method based privacy preservation model is proposed to modify the sensitive/quasi attribute sensitive values in a given dataset. The goal is to generalize the sensitive data rather than sanitization of sensitive/quasi attribute. Hierarchical clustering is used to split the data into clusters with objective of managing data chunks. For each sensitive attribute in a given cluster, attribute frequency distribution is formed to find the probability of values ranging from the lower-class limit to the upper-class limit. Median is chosen as a threshold to declare sensitive data in a given cluster. The sensitive numeric data is modified using S-shape membership function and sensitive categorical data is generalized by anonymization operation. At the end, sensitive and non-sensitive data is combined for mining operations. Data mining techniques can classify, cluster or make a decision tree without disclosing the individual information. For empirical analysis Bank Marketing and Adults datasets are used.

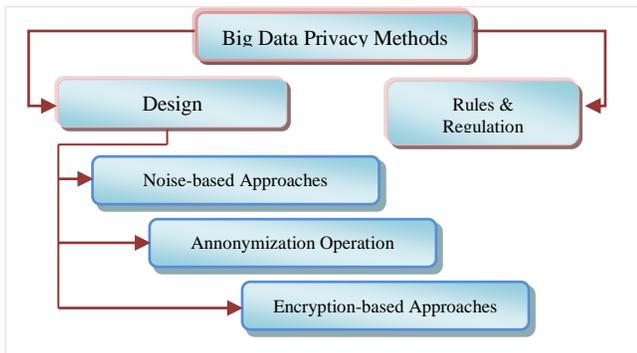

Fig. 1. Big Data Privacy Approaches.

Remaining part of paper is organized as follows; Section II contains related work done by previous researchers, in Section III we proposed our methodology, Section IV contains experimental section and Section V presents conclusion.

## II. LITERATURE REVIEW

This section review the earlier studies by highlighting the opportunities of data analysis for miners as well as the challenges faced by the researchers. The major challenges are storage, processing and privacy & security of Big data. First, the rapid origination of large volume of electronic data causes the incremental requirement of storage. Second, Big data is collection of different types of data such as structured, semi-structured and unstructured that make complicated the mining process. Lastly, security and privacy refer to keep it safe from attackers and preventing Big data in processing phase while using data mining techniques for knowledge discovery. In current era, privacy of an individual and organizational data in Banks and Hospital is a major challenge for their sensitive and quasi information [13, 14, 15]. The mining process is a threat to disclose the identity of an individual using their sensitive attributes like salary, Balance and disease. Using quasi attributes such as zipcode, sex and DOB, 85% to 87% individuals of U.S. residents can be identified by Joining quasi attributes with external/publicly available voter list dataset can be misused after the disclosure of privacy of an individual [10].

Another easier target, that reveal sensitive information, is health care data. In 2009, 120 million health care records were manipulated. Theft resource center reported 1190 breach cases of health care data between 2005 to 2014. About 307000 patient's records were inadvertently placed accessible on internet containing sensitive information such as patient name, DOB and phone number. In the first three months of 2015, 91 million health care records were hacked. These mentioned cases increased the demand of Big data analytics [9]. To preserve privacy, perturbation algorithm is proposed by Adebayo et al. [8] that removes sensitive attributes like gender, race, religious views and political views from the data. The removal of attributes refers to the sanitization. The use sanitization process almost eliminates the reconstruction ability of sensitive data over a large scale. [11, 12] considered min-max normalization, fuzzy logic, rail-fence and map range to mask the sensitive attributes to preserves the privacy of an individual. The use of fuzzy logic is not only enhanced the efficiency of clustering, but also reduced the number of passes and processing time. Min-max normalization transformed numeric data into specific range while rail fence is more suitable to mask categorical data and map range is appropriate on numeric data only.

Besides this, an augmented anonymization method with better privacy performance is proposed by Rajalakshmi et al. [16]. Data is transformed into sub-clusters using isometric transformation. Quasi identifiers are modified to hide the association between individual and its actual record. To the best of our knowledge, data anonymization is considered as a simple procedure with better privacy preservation of data. However, numerical attributes can only be preserved by [16] and supports similarity attack upto some extent. Another data-leak detection (DLD) is presented by Xiaokui et al. [17] to solve the privacy problem of sensitive information. For this purpose, fuzzy finger print technique is used data-leak detection model to figure out, design and implement with an aim to enhance the privacy during data-leak detection operations. On the other hand, fuzzy logic is used to protect privacy of data by transforming numerical attributes into linguistic information. Sensitive data is made public with fuzzy draft rate. Fuzzy based experimental results were better than k-anonymity method in context of loss and performance. The performance and information loss are minimized 48% to 59% and 40% to 50% respectively [18].

## III. PROBABILISTIC MODEL FOR PRIVACY PRESERVATION IN BIG DATA

In this section, the generalized view of our probabilistic Model for privacy preservation in Big data is presented as shown in Fig. 2. Big data analysis computationally discover trends, patterns and associations related to human relations and actions according to their behavior. Therefore, trends, patterns and association visualization are hard to show with in a table or graphically. In this perspective, clustering is the best solution to split Big data into manageable groups. These clusters contain the sensitive/Quasi and non-sensitive attributes. Assuming sensitive/quasi attributes in each cluster as given , a frequency/probability distribution is obtained on quasi numeric attribute (e.g. age) data by finding the probabilities for the lower-class limit to upper class limit. In case of categorical attributes, frequent data values are considered as sensitive for sensitive/quasi attributes with an ANDing operation on the entire set of sensitive attributes. The median of these probabilities is used as threshold to find sensitive and quasi attributes values with an aim to apply S-shaped fuzzy membership function and anonymization operation to preserve the privacy of the data.

### A. Big Dataset

A dataset ranging from tera bytes to peta bytes or exa bytes is considered as Big data. The origination of Big data on daily basis is increasing due to posts on social media like twitter and facebook, and digital pictures and videos on YouTube, mobile computing, file hosting and transactional records. There are four types of attributes in Big datasets [19].

- Identifier: An identifier is used to identify an individual uniquely such as CNIC_No, Social_Security_number and Employee_No.

- Quasi Attribute: Attributes that identify an individual by linking data with external data are known as quasi attributes. Quasi attributes can be categorized into two types and shown in Fig. 3.

    o Quasi numeric Attributes: Attributes based on numeric values such as age, zipcode and Date_of_Birth.

    o Quasi Categorical Attributes: Attributes consisted of textual information or combination of characters, for example sex.

- Sensitive Attributes: Sensitive attributes provide an insight of an individual's private and confidential data. These attributes can also have classified into two types.

    o Sensitive Numeric Attributes: Attributes with numeric values such as account_balance, salary and income.

    o Sensitive Categorical Attributes: Textual values-based attributes or combination of characters are known as sensitive categorical attributes. Examples are race, occupation and disease.

- Non-Sensitive Attributes: Attributes that does not help in revealing the privacy of an individual.

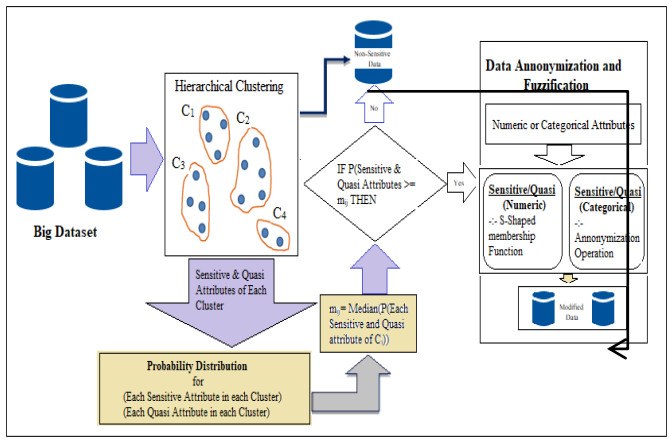

Fig. 2. Probabilistic Model for Privacy Preservation of Big Data.

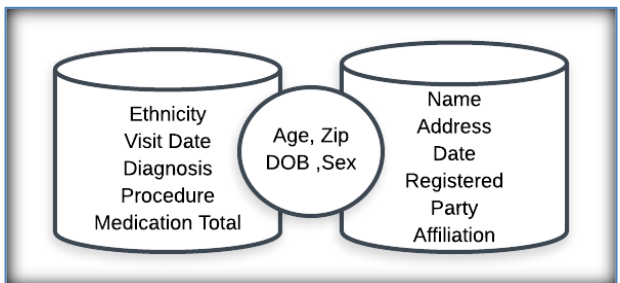

Fig. 3. Quasi Identifier.

## B. Clustering

Clustering is an unsupervised learning to group similar data of sensitive attributes. The goal of clustering is to find the fundamental groups in a set of unlabelled Big datasets, for example healthcare and bank marketting dataset. Data clustering can be used in various research domains such as data mining, pattern recognition, spatial databases, DNA analysis, Market analysis, medical domain and web statistics [20,21, 22]. Hierarchical clustering algorithm constructs a hierarchy among sensitive attributes values in order to split them into two or more clusters either on the basis of closeness or using a reversible process to merge clusters attributes values [23]. The outcome of the hierarchical can be represented by dendrogram [24, 25, 26]. For N sensitive attributes to be clustered, the basic process of $N$x$N$ distance similarity matrix of hierarchical clustering is as below:

> Assign each sensitive attribute data values to cluster based on similarities.
> Find the closet pair of cluster to merge clusters.
> Compute similarities b/w new & old clusters.
> Repeat step-2 & 3 untill all sensitive attributes are clustered.

## C. Probability Distribution of Sensitive Attributes

After clustering sensitive attributes, a probability distribution is obtained using the frequency of all sensitive and quasi attributes in each cluster. The probability of all sensitive and quasi attributes is computed according to the *age* attribute. This attribute is used to define the class ranges and the occurrences of *age* values in in each of these classes resulted in frequency of different people. The probability of each class can be obtained by equation 1.

$$P(A_{ij}(v)) = \frac{\left| A_{ij}^{c_i}(v) \right|_{lcl}^{ucl}}{|c_i|} -$$

(1)

Where $A_{ij}(v)$ represent the sensitive/quasi attribute of $i^{th}$ cluster of $j^{th}$ attribute's $v^{th}$ value, $c_i$ is the cluster number and $lc$ $l$& $ucl$ represent the lower-class limit and upper-class limit.

Median of $P(A_{ij}(v))$ is computed by equation 2 as a threshold to determine the sensitive/quasi attribute information that should be private before application of data mining technique. The probability above or equal to *M* is private data.

$$M = median(P(A_{ij}(v))) -$$

(2)

For the remaining sensitive attributes, ANDing operation is applied after thresholding (as given by equation 2) to consider sensitive/quasi attribute as private either by modifying it through S-shaped membership function or by anonymization operation.

## D. Fuzzification and Anonymization Operation on Sensitive Attributes Values

Fuzzy logic is considered as a remarkable method for data distortion with minimum information loss. Data distortion is the process of hiding sensitive attributes values without information loss. In this perspective, fuzzy transformation method is used to distort the sensitive/quasi attributes values by using the fuzzy membership function. The process of generating membership values for a fuzzy variable using membership function is termed as fuzzification [27, 28]. A membership function defines the mapping of membership value of input space between 0 and 1. The mapping of membership values of each sensitive/quasi attributes values (input space) can computed by equation 3.

$$S(\alpha,\beta,\gamma) = \begin{cases} 0, & \alpha \le \beta \\ 2\left(\dfrac{\alpha-\beta}{\gamma-\beta}\right)^2, & \beta \le \alpha \le \dfrac{\beta+\gamma}{2} \\ 1-2\left(\dfrac{\alpha-\gamma}{\gamma-\beta}\right)^2, & \dfrac{\beta+\gamma}{2} \le \alpha \le \gamma \\ 1, & \alpha \ge \gamma \end{cases}$$

(3)

Where $\alpha$ is the original data, $\beta$ represents minimum value in data set and $\gamma$ represents maximum value in data set. S shaped fuzzy membership function transforms the numeric sensitive information in each cluster into fuzzified form. Transformed data range between 0.0 and 1.0 [30].

For sensitive/quasi categorical attributes, anonymization operation replaces, modify and generalize the informative attributes values. The objective of anonymization operation is not to disclose or re-identify the sensitive/quasi categorical attributes data [31]. Anonymization operation can generalize, suppress, anatomization, permutation operation and perturbation operation [32]. First, generalization operation replaces sensitive and quasi attributes with some less specific values. For example, *age* can be generalized into ranges. Second, suppression operation uses special value (e.g. an asterisk '*') to replace sensitive/quasi attributes values. Third, anatomization dissociate the correlation observed between quasi and sensitive attributes data. Fourth, permutation operation shuffle values of sensitive/quasi attributes by partitioning records into groups. Lastly, Perturbation operation modifies the original data by transforming values with synthetic values.

### E. Algorithm for Modified and Generalized view of Data

The above-mentioned steps can be turned into algorithmic steps with an aim to modify and generalize the informative data of sensitive/quasi attributes. The important steps of the proposed algorithm are presented.

With the use of above algorithm, input dataset is modified only for sensitive and quasi attributes record with an aim to the minimal effect to the original data. Sensitive/quasi attributes modified using fuzzification through S-shaped membership function can be re-constructed. However, re-construction is out of scope of the algorithm. In addition, Sensitive/quasi attributes modified through anonymization operation cannot be re-constructed. The reason is that the sensitive/quasi attributes records are generalized. In this way, our proposed probabilistic model preserves the privacy of sensitive information of an individual before outsourcing the actual dataset.

```
1. Input D // Dataset
2. Set C_i = HierarchicalClust(D_i)
3. Assume Sensitive(S), Quasi Attributes(Q) & Non−Sensitive Attributes(N)
     S = {A_1, A_2, A_3, ..., A_n}, Q = {B_1, B_2, B_3, ..., B_n} & N = {I_1, I_2, I_3, ..., I_n} in each C_i
4. FOR each C_i
5.     IF A_i ∈ S THEN
6.         Compute P(A_i) in each C_i
7.     IF B_i ∈ Q THEN
8.         Compute P(B_i) in each C_i
9.     m_s = median(P(A_i(v)))
10.    m_q = median(P(B_i(v)))
11.    IF type(A_i) && type(B_i) = False THEN // Numeric
12.        IF P(A_i(v)) ≥ m_s AND P(B_i(v)) ≥ m_q THEN // Numeric (Sensitive / Quasi)
13.            DA_i = [A_i,1],[B_i,1] & Apply S−Shape Membership Function // Equ.3
14.        ELSE
15.            RA_i = [A_i,0],[B_i,0]
16.    ELSE   // Categorical
17.        IF P(A_i(v)) ≥ m_s AND P(B_i(v)) ≥ m_q THEN // Categorical (Sensitive / Quasi)
18.            MA_i = [A_i,1],[B_i,1] & Apply Anonymization operation
19.        ELSE
20.            UA_i = [A_i,0],[B_i,0]
21.    SC_i = FD_2 = [DA_i; RA_i], AD_i = [MA_i; UA_i] // Fuzzified, Anonymized and Non−Sensitive Records
22.    DC_i = concat(SC_i, N)
23. END
```

## IV. EXPERIMENTAL RESULTS

In this section, experimental results are discussed to demonstrate the cluster wise privacy of numeric and categorical attributes. Numeric and categorical attributes in each cluster can be divided into direct identifier, sensitive attributes and quasi attributes. Cluster wise probability-based threshold (e.g. median) is calculated for sensitive and quasi attributes. Numeric sensitive and quasi attributes having a median greater than probability is assumed as sensitive need to modify. Sensitive numeric information in the data set is modified by using S-shaped fuzzy membership and sensitive categorical information is transformed by using anonymization operation. After modifying all the sensitive numeric and categorical information of each cluster, probabilistic based model creates modified view of data without any loss of data and leak of privacy any one can use modified data for analysis purpose freely without any privacy issue. To test our proposed probabilistic based model, we considered UCI machine learning adult's dataset and bank marketing data set [31] for validating the results in comparison with sanitization. Both dataset information is presented in Table I.

To protect health related information in United States, HIPAA (Health Insurance Portability and Accountability Act) defines 18 elements that can be removed or generalized for ensuring privacy of data. Besides this, several sensitive and quasi attributes are commonly existing in most of the dataset that can reveal the privacy of individual. Therefore, sensitive and quasi attributes information should be private. Direct identifiers of HIPAA, Sensitive and Quasi attributes are presented in Table II.

TABLE I. ATTRIBUTES TYPES IN ADULTS AND BANK MARKETING DATASET

| Data Set | Instances | Attributes | Attrb. Types | No. Attr. Types | Sensitive Attributes | Quasi Attributes |
|---|---|---|---|---|---|---|
| Adults | 32561 | 15 | Identifiers | 0 | Race, Income, Occupation | Age, Sex |
| | | | Sensitive | 3 | | |
| | | | Quasi | 2 | | |
| | | | Non-Sensitive | 10 | | |
| Bank Marketing | 4521 | 17 | Identifiers | 1 | Balance, Loan, Job | Age |
| | | | Sensitive | 3 | | |
| | | | Quasi | 1 | | |
| | | | Non-Sensitive | 12 | | |

TABLE II. DIRECT IDENTIFIERS OF HIPAA, SENSITIVE AND QUASI ATTRIBUTES

| Direct Identifiers of HIPAA | Sensitive Attributes | Quasi Attributes |
|---|---|---|
| Invidual Name Individual Address(including street address, city county, and zip code) Dates related to an individual (including birthdate, admission date, discharge date, date of death, and exact age if over 89) Phone# Fax# Email_ID Social Security # (SSN) Medical record # Account # Certificate/licence # Vehicle/device serial# Web URL Internet Protocol (IP) Address Finger/voice print Photographic image | Salary, Disease Balance, loan status Race, Occupation Religion, income | Age Sex DOB Zip code |

### A. Clusters based Privacy of Adults and Bank Marketing Data

To split Big dataset into small and manageable fragments, hierarchical clustering is used to get clusters as shown in Fig. 4. Five clusters and three clusters were obtained from Adults and Bank Marketing dataset respectively after performing extensive experiments.

For each cluster, S-shape membership function and anonymization operation is used to secure the sensitive information according to the attributes type. These operations only modify the sensitive values which satisfy the probabilistic median based threshold criteria.

### B. Privacy through S-Shape Membership Function

S-shape fuzzy membership function, as given by eq. 3, distorts the sensitive/quasi numeric attributes by changing their values with re-construction ability. The reason is that S-shape fuzzy membership function mapps values in a range from 0 to 1 for each cluster. For example, age and balance attributes values are sensitive if threshold criterion is satisfied. Therefore, sensitive/quasi numeric values are modified by S-shape fuzzy membership function as shown in Table III.

The fuzzified view of Table III for sensitive attributes can be used for further analysis without breaching the individual privacy. However, the re-construction of both attributes (i.e. *Agea* and *Balance)* is out of scope.

### C. Privacy through Anonymization Operation

Anonymization operation replaces, modify and generalize individual data with an aim not either to disclose or re identified. For restricting disclosure or re-identification of data, data is anonymized before releasing it to the users.

Our proposed probabilistic model for privacy of data can handle categorical and numerical attributes for data mining by generalizing the attributes values. Data anonymization is the simplest procedure with better privacy. The application of anonymization operations can either be on categorical or quasi categorical attributes such as race, sex and occupation. The use of S-shaped fuzzy membership and anonymization operation ensures the best modification of direct, sensitive and quasi attributes by generalizing and fuzzified view of data for users. Table IV and Table V present the before and after modified view of data.

Similarly, S-shaped membership function and anonymization can also be used to ensure Big data privacy by modifying sensitive and quasi attributes values. Table VI and Table VII presents the results obtained on Bank Marketing dataset before and after modification for a better comparison.

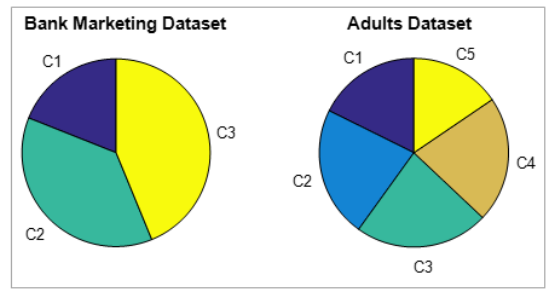

Fig. 4. Clusters of Big Datasets.

TABLE III. BEFORE AND AFTER MODIFICATION OF ATTRIBUTES VALUES BY S-SHAPE FUZZY MEMBERSHIP FUNCTION

| Before modification | | After modification | |
|---|---|---|---|
| Age | Balance | Age | Balance |
| 50 | 3143 | 0.817 | 0.00687 |
| 40 | 2096 | 0.397 | 0.01581 |
| 26 | 12519 | 0.061 | 0.03941 |
| 28 | 11262 | 0.091 | 0.88158 |
| 43 | 343 | 0.507 | 0.84953 |

TABLE. IV.    SAMPLE OF ADULTS DATA SET BEFORE MODIFICATION

| Cluster ID | Age | Marital Status | Ocupation | Race | Sex | Income |
|---|---|---|---|---|---|---|
| 1 | 50 | Divorced | Other-service | Asian-Pac-Islander | M | <=50K |
| 1 | 19 | Never-married | Sales | White | M | <=50K |
| 2 | 61 | Married-civ-spouse | Prof-specialty | White | M | <=50K |
| 2 | 40 | Married-civ-spouse | Craft-repair | White | M | <=50K |
| 3 | 18 | Never-married | Other-service | Black | M | <=50K |
| 3 | 26 | Married-civ-spouse | Prof-specialty | White | M | <=50K |
| 4 | 28 | Married-civ-spouse | Sales | White | M | <=50K |
| 4 | 71 | Married-civ-spouse | Farming-fishing | Amer-Indian-Eskimo | M | <=50K |
| 5 | 73 | Married-civ-spouse | Sales | White | M | >50K |

TABLE. V.    SAMPLE OF ADULTS DATA SET AFTER MODIFICATION

| Cluster ID | Age | Marital Status | Ocupation | Race | Sex | Income |
|---|---|---|---|---|---|---|
| 1 | 0.817 | Divorced | Unkonwn | Unkonwn | Person | Unkonwn |
| 1 | 19 | Never-married | Sales | White | Male | <=50K |
| 2 | 61 | Married-civ-spouse | Prof-specialty | White | Male | <=50K |
| 2 | 0.397 | Married-civ-spouse | Unkonwn | Unkonwn | Person | Unkonwn |
| 3 | 18 | Never-married | Other-service | Black | Male | <=50K |
| 3 | 0.061 | Married-civ-spouse | Unkonwn | Unkonwn | Person | Unkonwn |
| 4 | 0.091 | Married-civ-spouse | Unkonwn | Unkonwn | Person | Unkonwn |
| 4 | 71 | Married-civ-spouse | Farming-fishing | Amer-Indian-Eskimo | Male | <=50K |
| 5 | 73 | Married-civ-spouse | Sales | White | Male | >50K |

TABLE. VI.    SAMPLE OF BANK MARKETING DATA SET BEFORE MODIFICATION

| Cluster ID | Age | Job | balance | Day | month | Loan |
|---|---|---|---|---|---|---|
| 1 | 33 | management | 3143 | 29 | Jun | No |
| 1 | 45 | management | 2096 | 21 | Nov | No |
| 1 | 47 | admin. | 1934 | 14 | May | Yes |
| 2 | 50 | blue-collar | 12519 | 17 | Apr | No |
| 2 | 60 | Technician | 11262 | 26 | Aug | No |
| 2 | 68 | Retired | 4189 | 14 | Jul | No |
| 3 | 26 | management | 63 | 28 | Jul | No |
| 3 | 32 | Services | 182 | 6 | May | No |

TABLE. VII.    SAMPLE OF BANK MARKETING DATA SET AFTER MODIFICATION

| Cluster ID | Age | Job | balance | Day | month | Loan |
|---|---|---|---|---|---|---|
| 1 | 0.155 | Unknown | 0.0069 | 29 | Jun | Unknown |
| 1 | 0.228 | Unkown | 0.0158 | 21 | Nov | Unknown |
| 1 | 45 | management | 2096 | 21 | Nov | Yes |
| 2 | 0.831 | blue-collar | 0.0394 | 17 | Apr | Unknown |
| 2 | 0.369 | Unknown | 0.8816 | 26 | Aug | Unknown |
| 2 | 68 | Retired | 4189 | 14 | Jul | No |
| 3 | 26 | management | 63 | 28 | Jul | No |
| 3 | 0.165 | Unknown | 0.7405 | 6 | May | Unknown |

## V. COMPARISON OF RESULTS WITH EXISTING APPROACHES

After modification of cluster wise attributes values, the entire data of all clusters for sensitive, quasi and non-sensitive attributes were combined for open access. In this combined open access data, anonymized attributes cannot be re-constructed due to generalization. However, fuzzified view of sensitive and quasi attributes can be re-constructed. Fig. 5 presents the percentage of sensitive attributes including the total number of records and total number of attributes in each dataset.

Similarly, Fig. 6 presents the cluster wise records, modified records percentage and a probabilistic threshold used for sensitive attributes. Fig. 6(a) and Fig. 6(b) present the modification through S-shaped membership function on Adults and Bank marketing datasets, respectively.

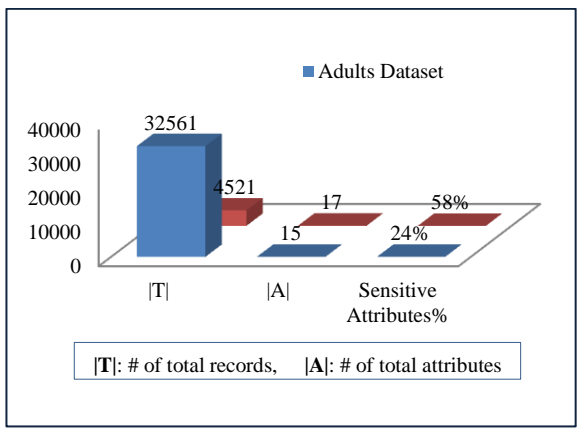

Fig. 5. Sensitive Information in Datasets.

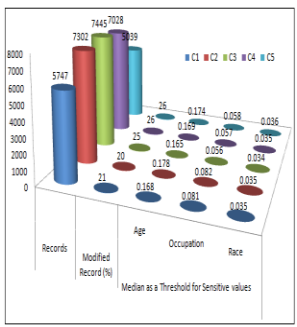

(a) Adults Dataset (Cluster Wise) Records, Modified view and Probabilistic Threshold.

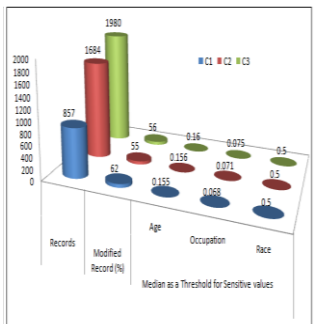

(b) Bank Marketing Dataset, Cluster Wise Records, Modified view and Probabilistic Threshold.

Fig. 6. Cluster Wise Modified View of Records.

Furthermore, our probabilistic threshold-based model for privacy preservation of big data is compared with the sanitization-based approaches [8] to validate the presented work. The objective of sanitization is to remove the sensitive attributes like gender, race, religious views, political view etc. from dataset. The problem with sanitization is left with only generalization and limits the application of data mining techniques for knowledge extraction. Sanitization restricts the re-construction of attributes upto 20% after removal of sensitive attributes. On the other hand, our probabilistic based threshold model can re-construct the fuzzification based modified data without removing any attribute. However, sanitization and our proposed model cannot re-construct the generalized attribute values. Thus, our model is more useful for users with the use of anonymization and fuzzification. In this perspective, our model modified only sensitive values of each cluster attributes upto 24% and 58% for adults and Bank Marketing dataset.

## VI. DISCUSSION

Social networking applications allow user to create content freely data which can cause a dramatic increase in data volume known as Big Data. Big Data have to be managed and analyzed efficiently to retrieve useful information to enhance business and improve society. Privacy preserving data mining emerged as a field to discuss privacy problem arise in data mining techniques to large scale individual data. Predictive Privacy harm itself is greater harm to individual privacy cause by sensitive attributes.

Sensitive attribute are part of mining process. Removal of sensitive attribute and quasi attribute causes data loses due to which big data is not useful for future data mining context. Browsing through a big data set would be difficult and time consuming to extract useful information from unstructured, invaluable, imperfect and complex data. Big data is divided into clusters using hierarchical clustering. Hierarchical clustering is more flexible and efficient approaches for clustering divide the data into different clusters using ward's method. Cluster wise probability and median of all sensitive and quasi attribute is calculated. Sensitive and quasi attributes whose probability is greater than median are declare as sensitive.

Probabilistic threshold based Big data privacy model creates fuzzified and anonymized view of sensitive values cluster wise using S shaped fuzzy membership function and anonymization operations. S shaped fuzzy membership function is applied upon numeric sensitive values enhance privacy of individual data by creating fuzzified view of sensitive information. Anonymization operation is performed upon sensitive values of categorical attributes and quasi identifier to preserve the privacy of individuals by creating anonymized view of big data. The Proposed model creates modified view of data by combining fuzzified, anonymized, non-sensitive attributes and values cluster wise. Modified Data can be used for analysis purpose. Modified data can be freely use for data mining purpose. The Proposed model reduce data lose and leak of privacy by creating modified view of data and modifying minimum no of attributes cluster wise.

## VII. CONCLUSION AND FUTURE WORK

Privacy preservation turns out to be an important aspect in Big data to restrict the disclosure of sensitive information before applying data mining techniques. In our proposed model, we tend to preserve the individual privacy in baig data.with zero or minimum side effects.Main goal is to uncover and conceal those sensitive items by Anonymization and fuzzification participate in exposing individual privacy to external world. The objective of the model is to preserve the individual privacy by anonymization and fuzzification for each cluster. Anonymization generalize the individual's data while fuzzification modifies the individual data with the re-construction capability unlike previous approaches. Different pervasive tests over case studies and different datasets are held to validate the effectiveness and accuracy of proposed work. After resulting data is obtained,it is assembled back as single cluster for data minig purposes. The experimental results have proven the better privacy over sanitization-based methods. However, re-construction of modified data is yet to consider as our future work. In future, k-mean clustering can be helpful in locating the most sensitive data based cluster with additional feature of finding more accurate sensitive item which can further be reconstructed to maintain the originality of big data.